\RequirePackage{fix-cm}
\documentclass[smallextended]{svjour3}       
\smartqed  
\usepackage{graphicx}
\usepackage{subfig}
\begin{document}

\title{BSROne: Binary Search with Routing of O(1); A Scalable Circular Design for Distributed Networks}



\author{Alireza~Naghizadeh\and Tahereh~Yourdkhani\and Behrooz~Razeghi\and Ehsan Meamari}

\authorrunning{A.~Naghizadeh, T.~Yourdkhani, B.~Razeghi, E. ~Meamari} 

\institute{A. Naghizadeh \at
				Department of Computer Engineering, Guilan University, Iran\\
              \email{alireza.naghizadeh.a@ieee.org}           
           \and
           T. Yourdkhani \at
           Department of Computer Engineering, Semnan University, Iran\\
             \email{ tahereh.yourdkhani@gmail.com}
            \and
           B. Razeghi \at
           Department of Electrical Engineering, Ferdowsi University of Mashhad, Iran\\
              \email{behrooz.razeghi.r@ieee.org}
            \and
           E. Meamari \at
           Department of Electrical Engineering, Iran University of Science and Technology, Iran\\
              \email{ehsanmeamari@gmail.com}   
}

\date{Received: date / Accepted: date}

\maketitle

\begin{abstract}
Peer-to-Peer (P2P) networks as distributed solutions are used in a variety of applications. Based on the type of routing for queries among their nodes, they are classified into three groups: structured, unstructured and small-world P2P networks. Each of these categories has its own applications and benefits. Structured networks by using Distributed Hash Tables (DHT) can forward request search queries more efficiently. These networks usually organize a specific topology and make a geometrical shape. A circular topology is a prevalent design which was first introduced by Chord. In this paper, we propose BSROne, a circular structured P2P design which attempts to consider several shortcomings in the current networks. In our proposed method, we want to achieve O(1) routing time without requiring all of the nodes to know about each other. By removing the real connections between nodes and tying all of them with super-nodes, we could reduce the number of overheads that are essential to maintain the connectivity between nodes in such networks. Furthermore, we gave the network an ability to scale up by introducing one layer above super-nodes. We achieved this by emulating the design of binary search algorithm for supreme-nodes. In this paper, at first we introduce a design where fixed super-nodes with unlimited resources are given to the distributed network. In the next step, we explain how it can manage to work as a P2P application. We finally discuss the possibility of removing the scalability issue in a P2P environment for our design.

\end{abstract}

\section{Introduction}
\label{intro}

Peer-to-Peer (P2P) networks are a subsidiary of distributed networks that are being used in a variety of applications \cite{lu2005stochastic}. These systems can be considered as an alternative of the client/server approach for routing between nodes \cite{naghizadeh2020gnm,hatamian2016cgc} and sharing resources \cite{shah2010cross}. They are also well-known for their two key characteristics, scalability and cost-efficiency \cite{naghizadeh2016improving,naghizadeh2015counter}. These factors led them to be a significant player for providing distributed services, in particular file-sharing. Therefore, routing for search queries plays an important role in the design of such networks.

There have been various approaches for designing such networks. Although, these methods can have different characteristics or follow different goals, but as a whole they can be categorized into three main groups: unstructured, structured and small-world. Structured P2P networks use Distributed Hash Tables (DHT) for routing. This means that peers should retain the information of the resources stored in other peers. The implementation of DHT tables can be done in various ways. But all these designs share the same purpose to direct queries as close to the destination as possible. In this way, they can provide a mechanism for finding the data objects within small amounts of messages in comparison to the total population of network \cite{punceva2003efficient}. 

The first problem in structured networks is how DHT tables update their entries. This problem originates from the fact that users join and leave these networks unexpectedly. As a result, keeping the track of these changes, leads to a lot of overloads in the network \cite{bhattacharya2011popularity}. Another major problem is the number of nodes which is needed for queries to reach their destinations \cite{yu2007bgkr}. In Multi-hop designs, we reach a successful lookup in O(log N) but in one-hop design we achieve this by O(1) \cite{kolberg2007markov}. Current O(1) networks usually have this problem that nodes should dedicate enough bandwidth and storage capacities to store information about all of the other nodes in the network \cite{buford2009p2p}. This is not a problem when network does not have a lot of nodes. But in the larger networks when scalability is required, since a lot of notifications should be sent in the network such systems will not be feasible  \cite{shen2010handbook}. 

In this paper, we propose Binary Search with Routing of O(1) (BSROne). We have two primary goals in mind for our network. At first it should have an O(1) routing time without hindering scalability of the network when it is required. We achieve this by using nodes with different status and emulating binary search methodology for scalability. As for our second goal we also intend to use a reactive recovery design for churn. This allows us to always have up-to-date tables when nodes join or leave the network. We use a reactive recovery while still reducing the amount of overheads compared to other methods which their primary purpose is to reducing overheads. 

This paper is organized as follows. Section 2 represents related work which is designed to make readers ready for our proposed method. First, we define different types of P2P networks and then in the second part we review previous work for the circular structured networks. The fundamental design of BSROne is discussed in section 3. In section 4, we improve the basic design of BSROne and make it ready for P2P environments. In section 5, we complete the design of our proposed method by introducing the scalable design for P2P environments. Finally, we analyze different parts of our proposed method with simulations in section 6.   

\section{Related work}
\label{sec:2}

\subsection{Infrastructures of P2P Networks}
One of the most important aspects of a P2P network is its overlay topology which can be divided into three groups: unstructured, structured and small-world \cite{han2010hybrid,naghizadeh2016c,naghizadeh2015preserving,naghizadeh2016structural}. In the following, we briefly demonstrate how each of these models works.

Unstructured P2P networks do not follow any specific order for making their overlays. In this scheme, there are no restrictions for either number of the files or number of the nodes in the network. This feature gives them a possibility to include as many nodes as possible. But as a downside, they depend on flooding algorithms to find the objects. Therefore, when the size of networks grows, these algorithms propagate a lot of messages on the network and hinder the scalability they were enjoying for unlimited restrictions \cite{chen2008enhancing}. In other words, the main problem of these networks is the high overloads that are created due to high amounts of unrelated query messages \cite{akbarinia2006reducing}.

We can use structured networks to apply more efficient search algorithms. Not only P2P structured networks support routing for both endpoints and nodes with a logarithmic routing time, but also they provide an opportunity for nodes to select their adjacent nodes according to optimization criteria \cite{chun2005impact}. Most importantly with their designs, such networks can guarantee an efficient search for all the files, even the scarce ones \cite{hui2004small}. To do this, the overlay usually follows a specific geometrical shape such as a circle or n-dimensional rectangular. Also, they apply a distributed hash table (DHT) which finds data in the network by using a consistent hashing function. There are several types of DHT tables. We can distinguish them by how nodes keep their adjacent tables to reach an efficient routing track among each other \cite{joung2007keyword}. 

Small-world is a phenomenon which can be seen in different places such as human relationships, computer networks, mailing systems etc. A network with such phenomenon should have two main properties, low hop numbers between every two randomly selected nodes and high clustering factor. There have been several suggestions to construct a graph with such properties. In most of them, connecting each node to several adjacent nodes and then maintaining a few links to some selected nodes which are random and far from them are essential for their designs \cite{hui2004small}. In such networks, the most challenging part is to design a routing algorithm which can make use of such phenomenon in case of occurrence.

\subsection{Circular P2P Networks}
To prepare readers for the next section, we present the other designs of circular structured networks. We tried to select from both one and multi-hop networks. These networks are as follows:

Kelips \cite{gupta2003kelips} follows a DHT overlay that applies extended memory and background overhead to obtain an efficient O(1) lookup. Its overlay is simple and different in construction and maintenance in comparison with existing systems. However, its lookup and churn mechanisms are extremely easy. They depend on advanced broadcast system among their members. Locality and load balancing are supported in this network. Similar to this goal is UnoHop which proposes an efficacious DHT algorithm with O(1) lookup implementation that allows balanced bandwidth at all the nodes with decreasing the propagation delays \cite{sitepu2008unohop,gupta2004efficient}.

OneHop is an O(1) overlay which is divided into different slices. Each slice consists of several units. Both slice and units have leaders to control churn events. In fact, OneHop creates unbalanced load on units and slice leaders. In \cite{wei2010research}, the authors study inequality features of nodes in the network which can categorize nodes with the same properties. 

Epichord \cite{leong2006epichord} is a modified Chord that aims to accelerate the lookup process. It can approach to O(1) lookup efficiency in high lookup traffic and at least O(logN)-hop lookup performance under high churn in the worst condition. D1HT \cite{monnerat2006d1ht} is a O(1) DHT that  claims to reach the maximum performance with suitable overhead even though the size of networks is large.  

Chord\# \cite{schutt2006structured,schutt2008range} is a modified version of Chord algorithm that decreases the expense of updating from O(log N) to O(1) and narrows the search on lookup performance. Unlike Chord, Chord\# does not apply constant hashing functions but uses a key-order preserving functions to produce object IDs. It creates asymmetric load in the network.  

PChord \cite{huang2010pchord} is a search mechanism in two directions and uses bidirectional routing tables, clockwise and counterclockwise. In addition, it improves the speed of accessing resources and distributing the messages in the network. Hybrid-Chord \cite{flocchini2004hybrid} is similar to PChord but it uses several Chord rings and successors to improve the routing performance. In addition, it can decrease the number of hops considerably. We put \cite{lu2009ml} in the same group. It is a P2P model, named ML-Chord, that allocates nodes into several layers like Chord according to classes of shared resources. This method provides both efficiency and scalability into consideration. A new version of this type of networks is \cite{naghizadeh2017binary}. 

Zhenhua Tan \cite{tan2008three} proposes a three layer protocol routing named CSSP. It improves routing speed with offering a smaller size of routing table and uses O(1) node finger to get O(N) routing path. In comparison with Chord, Pastry etc, CSSP improves the maintenance of routing table, routing hops, fault-tolerance while churn happens. 


\section{Fundamental design of BSROne}
\label{sec:3}
There are two problems in the current infrastructures which we want to resolve in our proposed method. In O(1) designs, each node should know about all of other nodes in the network. The obvious obstacle for this design is scalability. Another issue which can be found in almost all the circular structured networks is how they update their DHT tables and connections. This is a critical issue when churn happens (nodes join or leave the network). The reason is that the network should maintain its connectivity and update its DHT tables in churn. There are usually two approaches which are taken by current networks; reactive recovery and periodic recovery. 

In reactive recovery when churn happens, an immediate response takes place to update the network. An example for a network which uses this method is Free Pastry. The main problem with this method is high overloads, especially for updating DHT tables. Another method which is common and can be seen in the networks such as Chord and its derivations is periodic recovery. In this case, instead of an immediate reaction, each node should update its connections and tables periodically without considering churn. In this case, even though there are less overloads in the network but it still has to deal with a lot of signals since each node has to send signals for updating its connections and tables. There is also a bigger problem that the network should struggle to reach for its ideal status. This means that tables or connections may not point to the nodes they are supposed to. For more information about the consequence of this method, we refer readers to \cite{liben2002analysis,rhea2004handling}.

In our proposed method, with its basic design we aim to 1- have an O(1) network without requiring for information about all the nodes in the network, 2- use reaction recovery without bringing a lot of overloads on the network. We achieve this by clustering the network and using the super-nodes instead of equal treatment of all nodes. 

\subsection{Basic Methodology}

To describe the basic design, we assume an ideal network where the super-nodes are provided for the network. The network is very similar to the other circular designs where nodes are sorted in ascending order and can join and leave the overlay frequently. In real world applications, networks should use hash algorithms to specify ID numbers. But here, we use real numbers in our figures for simplicity. Like Chord, each network can have a maximum number of $2^n$ nodes.

The main difference in our design compared to other networks begins with connectivity. For network in order to stay connected, we get help from super-nodes. This works in contrast to the other networks which link each node to their predecessors and successors directly.

\begin{figure}[!h]
\centering
\includegraphics[width=72mm]{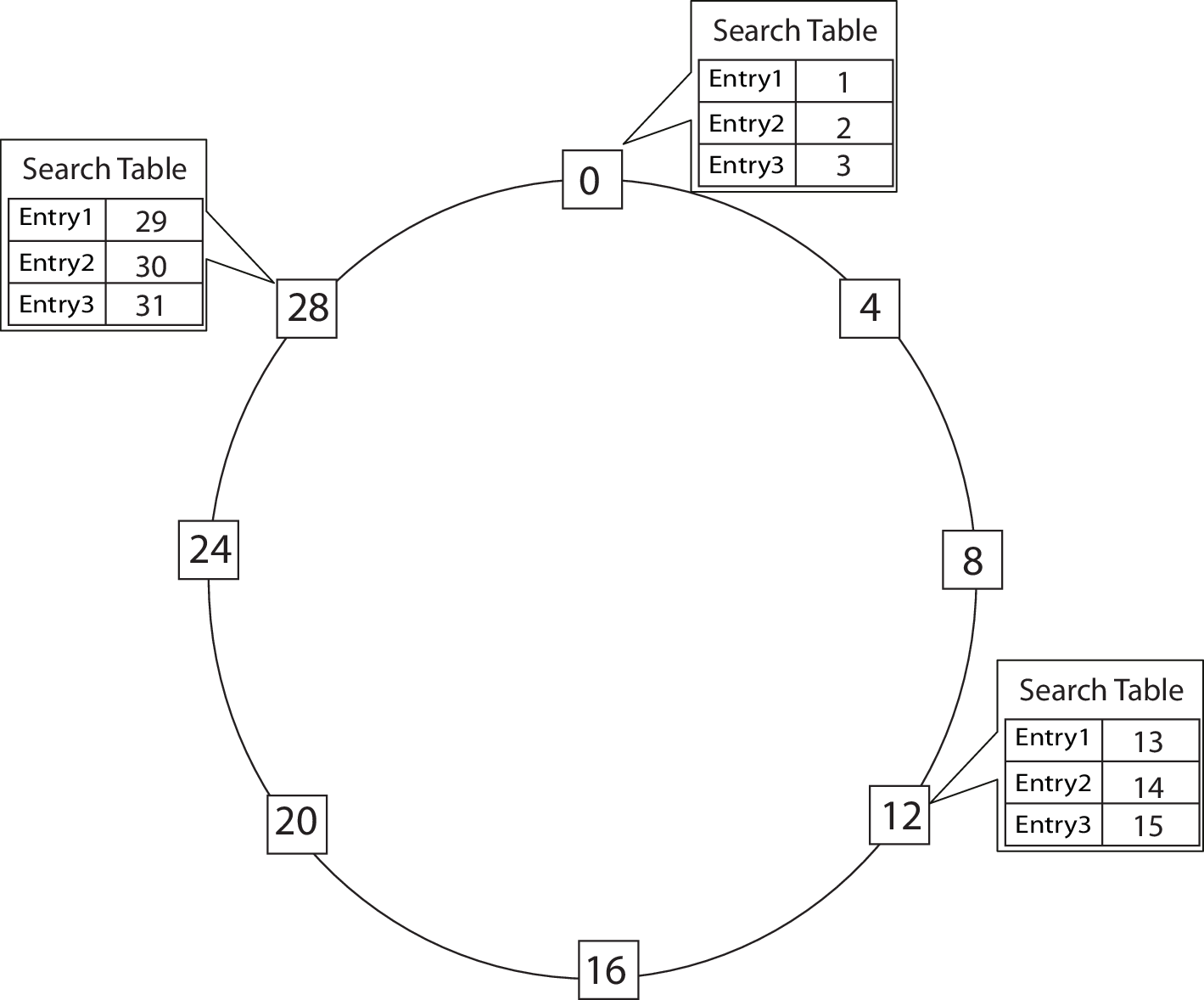}
\caption{An example of search tables in BSROne.}
\label{fig_1}
\end{figure}  

Another major difference is that in BSROne, unlike its counterparts, routing and searching are two different jobs. Therefore, we consider two types of tables named search and routing tables for each super-node. The super-nodes can find the other nodes locally by search tables. If the node was not available in the cluster then the routing tables are applied for jumps. 

\begin{figure}[!h]
\centering
\includegraphics[width=72mm]{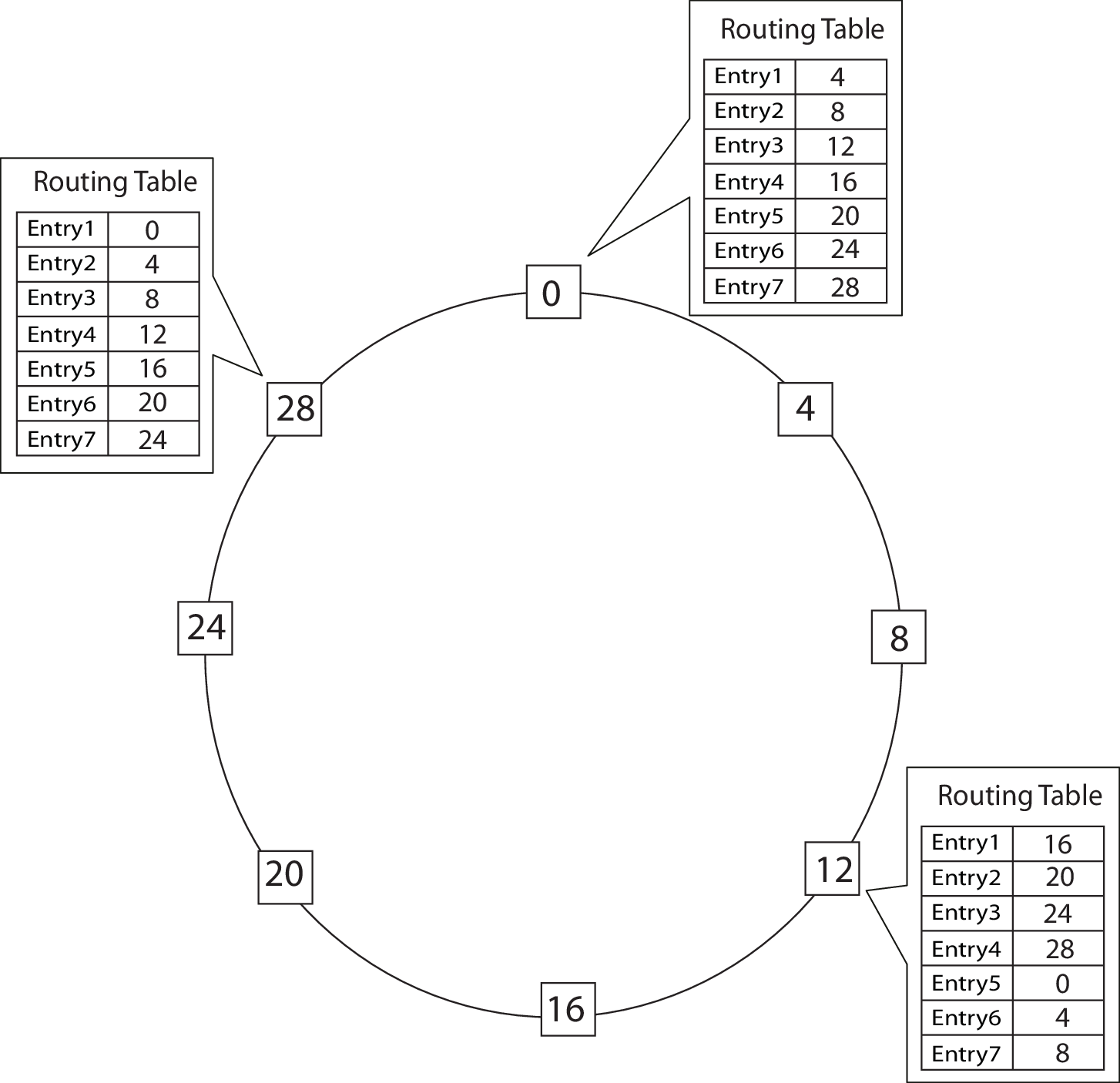}
\caption{An example of routing tables in BSROne.}
\label{fig_2}
\end{figure}

We can see the basic form of search tables in Fig.~\ref{fig_1}. As it is demonstrated, each super-node has one search table which can look for its cluster with clockwise order. The size of search tables should be proportional to the maximum size of the network. With a network of at most $2^n$ nodes, each search table can have at most $2^x-1$ entries where $2^n \% 2^x =0$. In this case, the network has $2^5=32$ nodes. The maximum number of nodes for each cluster is $2^2=4$ so search tables have $2^2-1=3$ numbers. 

In addition to search tables, we also need routing tables when requests are outside the scope of current clusters. Routing does not take place in the default state of BSROne. In this scheme, each super-node has to maintain a routing table containing information about all of the super-nodes in overlay. The super-nodes, therefore, can reach by one move to the other clusters of the network.

As we can see in Fig.~\ref{fig_2}, a network with 7 clusters is presented with 3 samples of routing tables. It shows that the number of entries in each table is equal to the number of clusters. Only one routing table is required for each super-node which is filled with the other super-nodes in clockwise order.

\subsection{Considering Scalibility}

In the prior section, we considered a default environment where each super-node was aware of other super-nodes in the network. Although, it is a very efficient design which gives us an O(1) routing time, it may be a barrier for scalability. To fix this problem, we define supreme-nodes which should coexist with current super-nodes. The supreme-nodes are the most powerful super-nodes in each section of the network which should locate at the beginning of each section. For routing, instead of limiting them to know about all other supreme-nodes we allow them to expand much further by following the philosophy of binary search in P2P networks. This means that in each step of routing, at least half of the useless nodes will be ignored. 

\begin{figure}[!h]
\centering
\includegraphics[width=72mm]{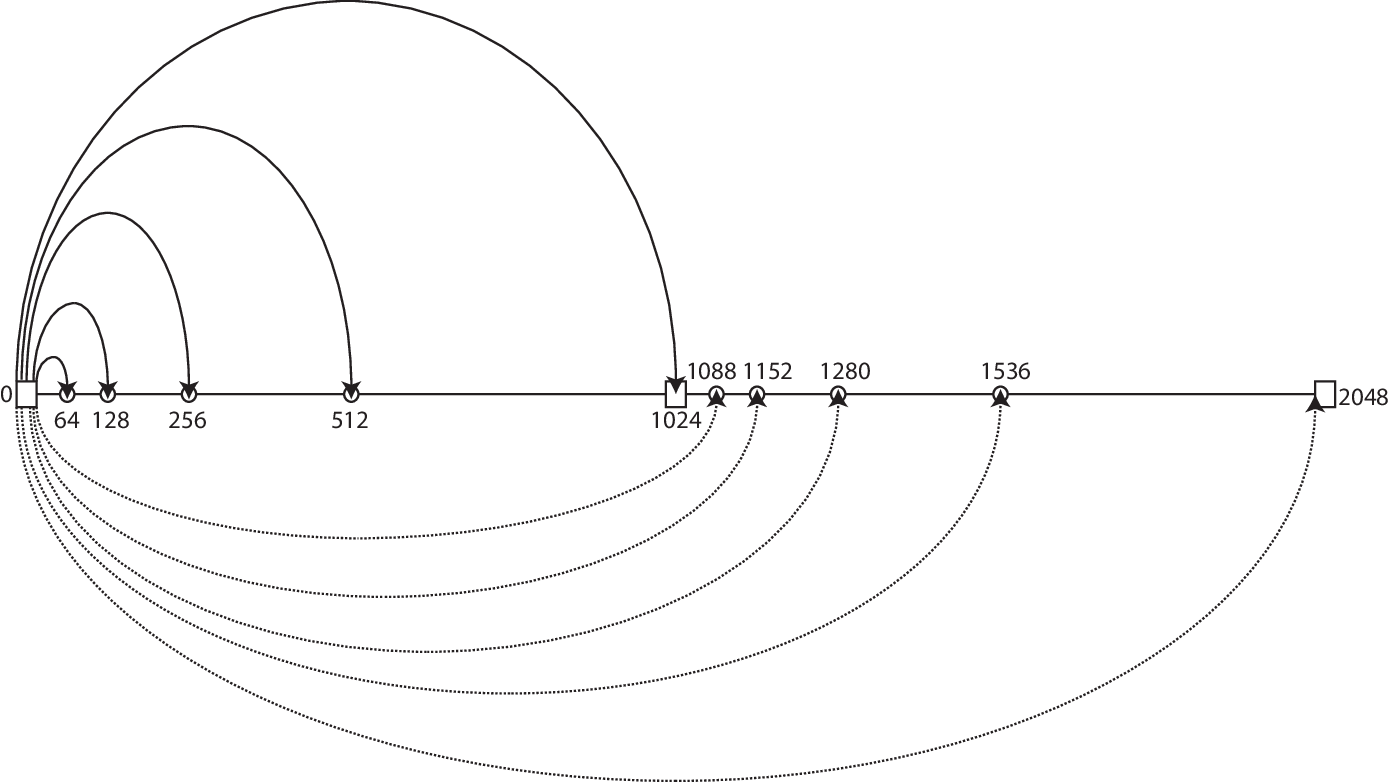}
\caption{An example of how we narrow our searches to the nearest point with routing tables.}
\label{fig_3}
\end{figure}

Assume that our network can have a maximum number of 2n nodes. Each supreme-node uses two routing tables instead of one. Each routing table can only see n, half of the maximum number of nodes in a network. Assume that our network is divided into smaller sections (these sections form smaller clusters themselves as it was seen before). Each section can have at most $2^x$ entries. The entries in clockwise routing table are filled as $2^x$, $2^{x+1}$, \dots , $2^{x+L}$ $where$ $2^{x+L} < n/2$, $n/2$, $n/2 + 2^x$, $n/2 + 2^{x+1}$, \dots ,$n/2 + 2^{x+L}$ 		 $where$ $n/2 + 2^{x+L} < n$, $n$. A similar pattern is also used for routing on the other half of the network for counterclockwise routing tables.

An example of a big network which is divided into smaller sections for every 32 nodes is depicted in Fig.~\ref{fig_3}. This shows how jumps take place with supreme-nodes. As it can be seen from the figure, not only we are able to halve the network, but also with this design we can guide our routing to the more nearest places. We will come back to this subject in section 5 and show how routing should take place when there are not fixed servers in the network.

\section{Improvement for P2P Environment}
\label{sec:4}

In the previous section, we assumed that there are fixed servers in the network which can give us required resources. But a P2P environment cannot usually benefit from such luxury. As a result, from now on, we assume that there is no server in the network. So we explore how regular nodes should replace super-nodes autonomously. 

\subsection{Management of Joining Nodes}
\label{sec:4.1}

\begin{figure}[!h]
\centering
\includegraphics[width=100mm]{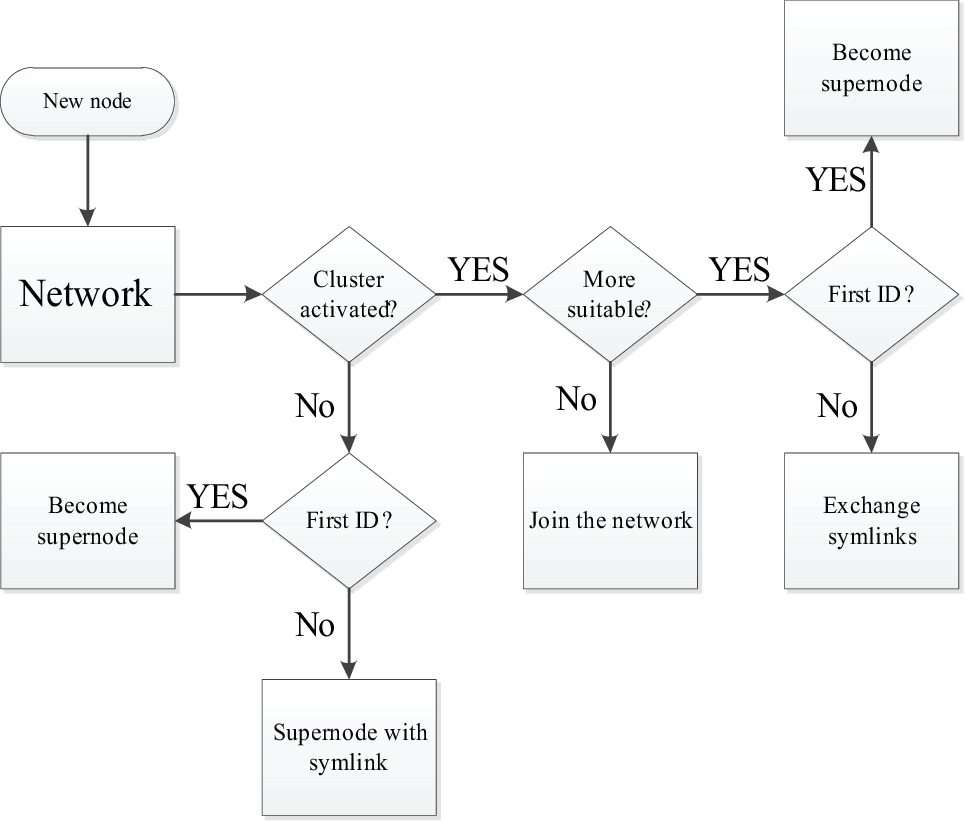}
\caption{A flowchart of how a newcomer joins the network.}
\label{fig_4}
\end{figure}

When there are no predefined servers to perform as super-nodes, we have to define how new nodes should fill the positions of super-nodes. To do this efficiently in an autonomous manner, several factors should be considered. The whole process is illustrated in Fig.~\ref{fig_4}. This section is dedicated to explain this flowchart.  

As we mentioned, each node in the network has a unique ID number and clustering of the network is determined before joining all the nodes. At first, all of these clusters are deactivated. When a new node joins the network based on what number it posses, it may activate one cluster in the network. This action takes place when there is no other node in the cluster. The type of activation depends on the ID numbers. The reason for this complication is that a super-node should be the first node in each cluster. This means that in a network with 32 nodes which is divided into 8 clusters, IDs of 0, 4, 8, \dots , 24, 28 should be filled first. But since nodes acquire their ID numbers randomly, this may not happen and a node takes the position of the super-nodes that does not posses one of the first IDs. To solve this problem, we use the concept of symlink. This means that a node can have two different IDs. One of them is a real ID and the other is used as a symlink to take only the job of super-nodes. With this approach, two results may occur. The node with the first ID in the cluster claims its job before anyone else or another node in that cluster takes the job. In the first situation, we do not have any problem. In the second case, it will be revealed for the other super-nodes that this ID is only a symlink and the real node with that number is not presented in the network.

The second part of the diagram is related to a situation that the cluster of that node is already created. In this case, we need to answer a different question. The first chosen nodes are not necessarily the best ones to act as the super-nodes. So we need a mechanism to exchange super-nodes when it is necessary. Therefore, when a new node joins the network and if it was not better than any other super-nodes in the network, it would simply join its cluster. But if it was revealed otherwise, we would need to position it as a super-node. The next step depends on the ID number of the node. If it is the first node of a cluster, then it simply becomes a super-node. But if it is not, then a symlink exchange should take place.

\begin{figure}[!h]
\centering
\includegraphics[width=72mm]{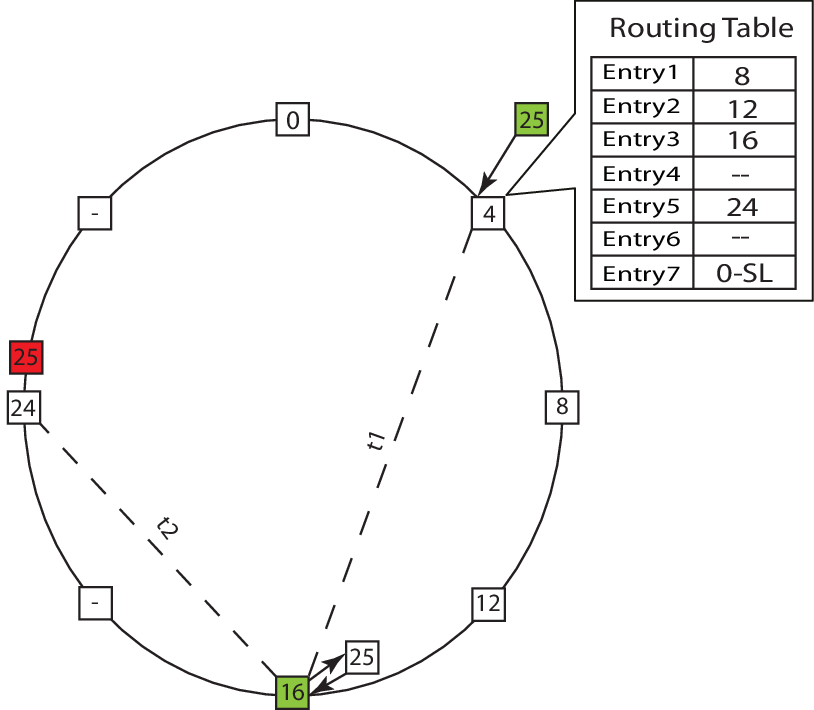}
\caption{An example of substitution of a newcomer.}
\label{fig_5}
\end{figure}

To clarify this process, we give an example that shows how a newcomer joins the network. In this example, we only show one scenario but this should clarify things about how other scenarios work too. This process can be seen in Fig.~\ref{fig_5}. As we can see, a network is presented with at most 32 nodes which are clustered into 8 sections. Clusters of 20 and 28 are not active and super-node 0 is marked as SL. The reason is that the real node of 0 is not presented in the network and another node took the job of super-node.  

As it is illustrated in the figure, node 25 wishes to enter the network. First, it sends a message to super-node 4. This super-node compares it with the other super-nodes in its routing table and realizes that node 25 is more suitable for super-node 16. So their ID numbers will be exchanged in t1 and the new node 25 goes to its own cluster in t2. After that, when a node contacts node 16 for its data, it will be forwarded to node 25 since the real data is in that node. The same process happens with node 25. When node 25 is requested for its real position, it forwards the request to node 16.

The only problem which remains is when a node joins the network and it does not appear to be suitable for the task of super-node at the beginning. But after a while, this is quite possible that it reaches to a point where it is more suitable than the other super-nodes for replacement. For this purpose, each super-node should keep track of the members in its cluster and receive their attributes regularly. If such thing happens, just like the example shown, it should exchange its ID with a super-node.

\subsection{Management of Leaving Nodes}
\label{sec:4.2}

The design of BSROne is very efficient for the leaving process when it occurs for regular nodes. All a super-node needs is changing its search table. But this is especially important when super-nodes themselves leave the network. In this situation, we may lose a cluster altogether.

\begin{figure}[!h]
\centering
\includegraphics[width=72mm]{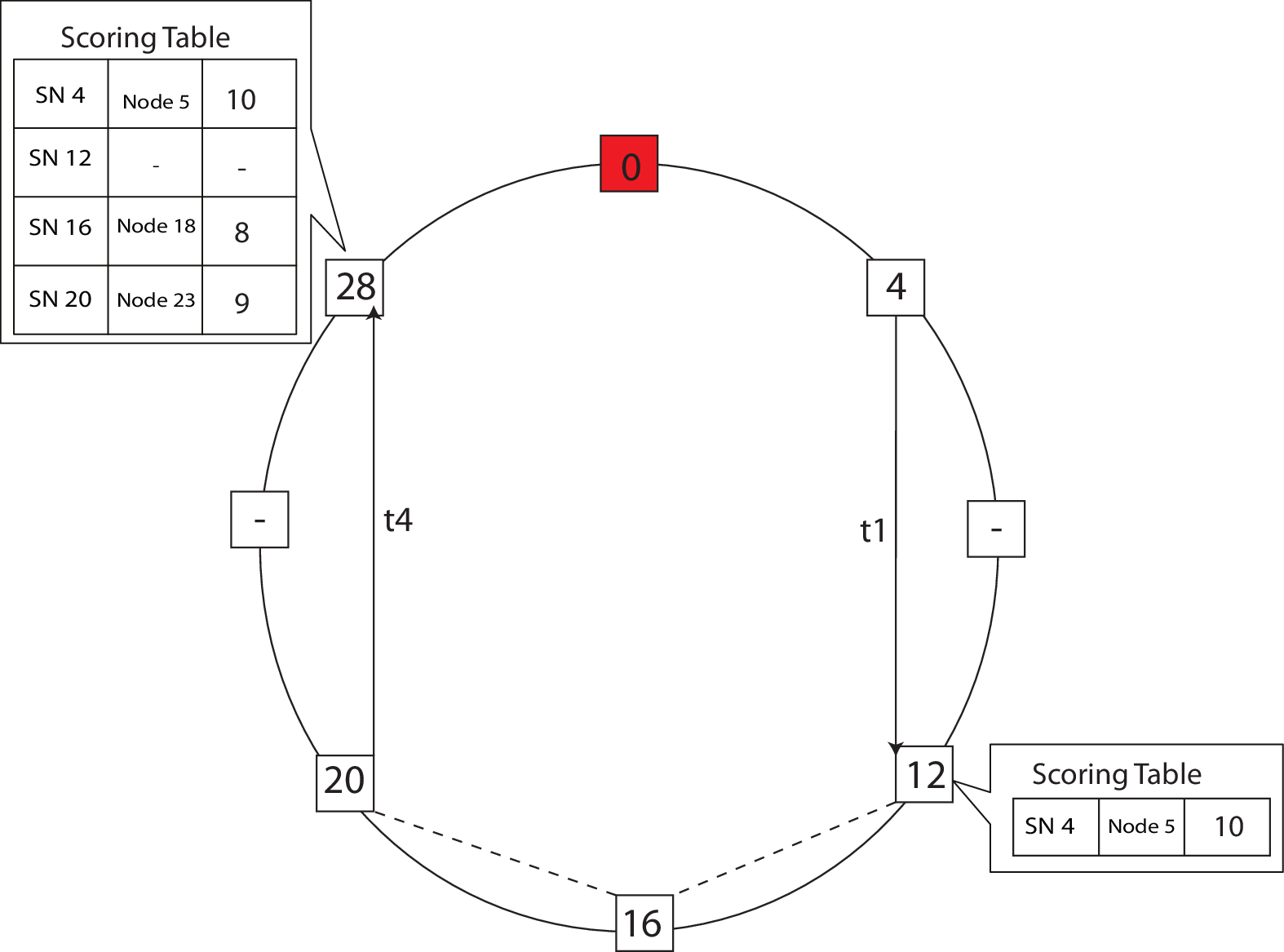}
\caption{The process of selecting a super-node after one has left the network.}
\label{fig_6}
\end{figure}

As previously mentioned, the idea sets out to give the best node in the network to act as a super-node. Therefore, when a super-node leaves the network, the strongest node should replace it. This mechanism is done by sending a message to all the super-nodes one by one. When the message passes through each super-node, they give a score to their strongest node and register it on a table and pass the message. Finally, the scores are compared and the most suitable node with the greatest score will be chosen. Fig.~\ref{fig_6} represents this mechanism.

In this network, node 0 has left the network so now it should be decided what node replaces its place. In time t1, node 4 sends a message for the next super-node. Since clusters of 8 and 24 are inactive, node 12 will be chosen as the next super-node. This message contains at least three values, the super-node number, its best node in the cluster and its score. Similarly, the process continues until it reaches node 28. As we can see in t4, all the super-nodes in the network participate in this process. The only super-node which does not offer any replacement is node 12. The reason is that in cluster 12 the super-node is the only node in the network. Therefore, nothing could be suggested.

Since this process takes some time, each cluster should be ready for such an event. This means that each super-node should always choose a substitute to replace its place temporarily. This substitute does not actively participate in management but only synchronizes the required tables until super-node leaves the network. Therefore, the chosen node will be given to the substitute in the end and if it was a more suitable node for replacement, they would exchange their ID numbers. In this way, the most suitable nodes remain as super-nodes even after departure.        

\subsection{Super-node Selection}
\label{sec:4.3}

In the previous section, we talked about the suitable nodes for taking the job of super-nodes. But it was not clarified which factors should be considered for making such decision. In this section, we propose a novel method for super-node selection based on desired attributes. In the prior networks, only one key attribute usually would take into account but in this attempt we incorporate several key factors . This allows us to have a much smarter network which can be formed autonomously. Other methods similar to the methodology used in \cite{razeghi2015novel} or unsupervised methods \cite{naghizadeh2018meaningful} can also be used for selection.

In our method, we jointly consider four attributes for process of super-node selection. The attributes are:
\begin{itemize}
\item Bandwidth - since super-nodes should manage requests from the nodes plus their regular downloads and uploads, it is logical to take the nodes with more bandwidths.   

\item Time on the network - it is important to consider how much a super-node tends to stay in the network. A node which has the best bandwidth but leaves the network in a short time may bring more harms than benefits. Different parameters can be taken into consideration for estimating this parameter. Such as 1- its stability in the network, 2- its requests for the download (bigger downloads means it stays longer), and 3- its previous records. Any parameter we choose, estimating and including this attribute is an important step for choosing a super-node. 

\item Number of ID exchanges - we want to prevent a situation where two nodes periodically change their roles as super and regular nodes. To help this process, we can consider the number of times a node has changed its ID numbers as an attribute.  

\item Tendency to cooperation - even though being a super-node is not a matter of choice in some situations, we still can give users a degree of free will. An option can be presented for users and ask them how much they are willing to cooperate.
\end{itemize}

Since we have to make decision for super-node selection based on the above criteria, we employ the optimization methods which are using in multiple criteria decision making. It is obvious that the above attributes have no equal importance in the process of super-node selection. So for taking into account the importance of each criterion, we weight them based on their priorities. This means that network designers have the ability to allocate their desired weight to these attributes. We define the weight of criteria $c$ as $w_{c} \in \left[0,1\right]$, such that
\begin{eqnarray}
\sum_{c=1}^{4} w_{c} = 1.
\end{eqnarray}
Now, we define criteria weighted vector as $\mathbf{c}_w= {\left[
\begin{array}{cccc}
w_{\mathrm{1}} & w_{\mathrm{2}} & w_{\mathrm{3}} & w_{\mathrm{4}}
\end{array} \right]}^t$. 

Our method for super-node selection is similar to the mathematical \emph{technique for the order preference by similarity to ideal solution} (TOPSIS) which was introduced by Hwang \& Yoon \cite{yoon1995multiple}. In our method, a network designer determines the desired value of mentioned criteria first, i.e., the upper and lower bound of values which each criterion should be achieved. We define the upper bound of ideal values of these criteria by   
\begin{eqnarray}
\mathbf{q}^+ = {\left[\begin{array}{cccc}
q_1^+ & \, q_2^+ &  \, q_3^+ &  \, q_4^+ \end{array}\right]}^t = {\left[\begin{array}{cccc}
B^{+} & \, T^{+} &  \, K^{+} &  \, W^{+} \end{array}\right]}^t,
\end{eqnarray}
and the lower bound of ideal values of them by
\begin{eqnarray}
\mathbf{q}^- = {\left[\begin{array}{cccc}
q_1^- & \,q_2^- &  \, q_3^- &  \, q_4^- \end{array}\right]}^t ={\left[\begin{array}{cccc}
B^{-} & \, T^{-} & \, K^{-} & \, W^{-} \end{array}\right]}^t,
\end{eqnarray}
where $B^{+} / B^{-}$ is the max/min bound of ideal value for super-node bandwidth, $T^{+} / T^{-}$ is the max/min bound of ideal time for presence of super-node in the network, $K^{+}/K^{-}$ is the max/min number of exchanging IDs in the network, and $W^{\left(.\right)}$ is the willingness cooperation degree of a node in the network. The willingness cooperation degree is a number in which each node determines its number in the network that these values are in the set $\left\{0, 1, ..., 10\right\}$. The number $0$ denotes that the node has no willingness to join the cooperation pool, while the number $10$ denotes the most willingness to join the cooperation pool. Although, some of the criteria may have no maximum limits for the network designer, all of these criteria have minimum expected values. For the criteria $T$ and $K$, a network designer considers the upper bound value that can be achieved in practice. 

Assume that we have $N$ candidate nodes for super-node selection process. So we define the decision matrix as follows
\begin{eqnarray}
\mathbf{D} = \left[ \begin{array}{cccc}
d_{11} & d_{12} & d_{13} & d_{14}\\
d_{21} & d_{22} & d_{23} & d_{24}\\
\vdots & \vdots & \vdots & \vdots \\
d_{N1} & d_{N2} & d_{N3} & d_{N4}\\
\end{array}\right]
\end{eqnarray}
where $d_{ij}$ denotes the value of candidate node $i$ in criterion $j$. The values of different criteria are in different measurement units. For example, the unit of entries of first column is in $\mathrm{b/s/Hz}$ while the values of entries of second column are in $\mathrm{s}$. Employing mathematical operations for the operands with different units is not possible. Therefore, we exclude the dimension of these values by normalizing the decision matrix $\mathbf{D}$. Here, we use Euclidean Normalization to normalize the matrix $\mathbf{D}$. The entries of new normalize matrix $\mathbf{A}$ are given by
\begin{eqnarray}
a_{ij} = \frac{d_{ij}}{\sum_{i=1}^{N} d_{ij}^{2}}, i = 1, .., N, j = 1, ..., 4 \, . 
\end{eqnarray} 
Now, the weighted normalized matrix $\mathbf{W}$ can be computed as
\begin{eqnarray}
\mathbf{W} = \mathbf{c}_w \mathbf{A},
\end{eqnarray}
and the normalized constraint vectors as
\begin{eqnarray}
\mathbf{q}_{\mathrm{norm}}^{+} &=& {\left[\begin{array}{cccc}
\frac{q_1^+}{q_1^+} & \, \frac{q_2^+}{q_2^+} &  \, \frac{q_3^+}{q_3^+} &  \, \frac{q_4^+}{q_4^+} \end{array}\right]}^t={\left[\begin{array}{cccc}
1 & \, 1 &  \, 1 &  \, 1 \end{array}\right]}^t, \nonumber \\
\mathbf{q}_{\mathrm{norm}}^{-} &=& {\left[\begin{array}{cccc}
\frac{q_1^-}{q_1^+} & \, \frac{q_2^-}{q_2^+} &  \, \frac{q_3^-}{q_3^+} &  \, \frac{q_4^-}{q_4^+} \end{array}\right]}^t = {\left[\begin{array}{cccc}
\hat{q}_1^- & \, \hat{q}_2^- &  \, \hat{q}_3^- &  \, \hat{q}_4^- \end{array}\right]}^t.
\end{eqnarray}

Next, we compute the Euclidean distance of each candidate node to the upper and lower bounds of desired constraints. In fact, each candidate node is a point in $4$-dimensional space which its distance to another point in that space vector represents its distance to another candidate node. So the distance of each candidate node $j$ to the upper bound values is given by
\begin{eqnarray}
E_{i}^{+} = \sqrt{\sum_{j=1}^{4}{\left(a_{ij}-1\right)}^2}, j= 1, ..., N
\end{eqnarray}
and the distance of each candidate node $j$ to lower bound values is computed as
\begin{eqnarray}
E_{i}^{-} = \sqrt{\sum_{j=1}^{4}{\left(a_{ij}-\hat{q}_i^- \right)}^2}, j= 1, ..., N
\end{eqnarray}

Finally, in order to choose the superior candidate node for super-node selection, we rank the nodes based on their proximity to upper bounds and their distance from lower bounds. So we define the priority criterion as
\begin{eqnarray}
C_i = \frac{E_{i}^{-}}{E_{i}^{-} + E_{i}^{+}}.
\end{eqnarray} 

The value of $C_i$ denotes the priority of nodes in super-node selection process, i.e., greater value represents higher priority of node.

\section{Scalable P2P Environment}
\label{sec:5}

In this section, we consider a scalable design of BSROne for a P2P environment. This design is very similar to the basic P2P design described in previous section. Just like before, bigger sections should remain inactive until at least one node joins the network. In a scalable mode it does not matter what cluster of the section is activated, the first thing to do is to activate the first node of the section. By doing this, the nodes in that section can join the whole network. In addition, since the first node should work with the other parts of the network apart from insider  clusters, it is obvious that the first super-node in each section should be the best super-node. Since not only do they act as a super-node for their clusters but also they relate that section to the other sections of network. 

Moreover, protecting a supreme-node in the case of departure is more delicate. The reason is that in this case not only a cluster but also the whole section that a supreme-node represents may lose its connectivity. The process for protecting these nodes is just like protecting super-nodes. But for the reasons we mentioned above it is advisable to reserve more backups for these nodes.

\begin{figure}[!h]
\centering
\includegraphics[width=100mm]{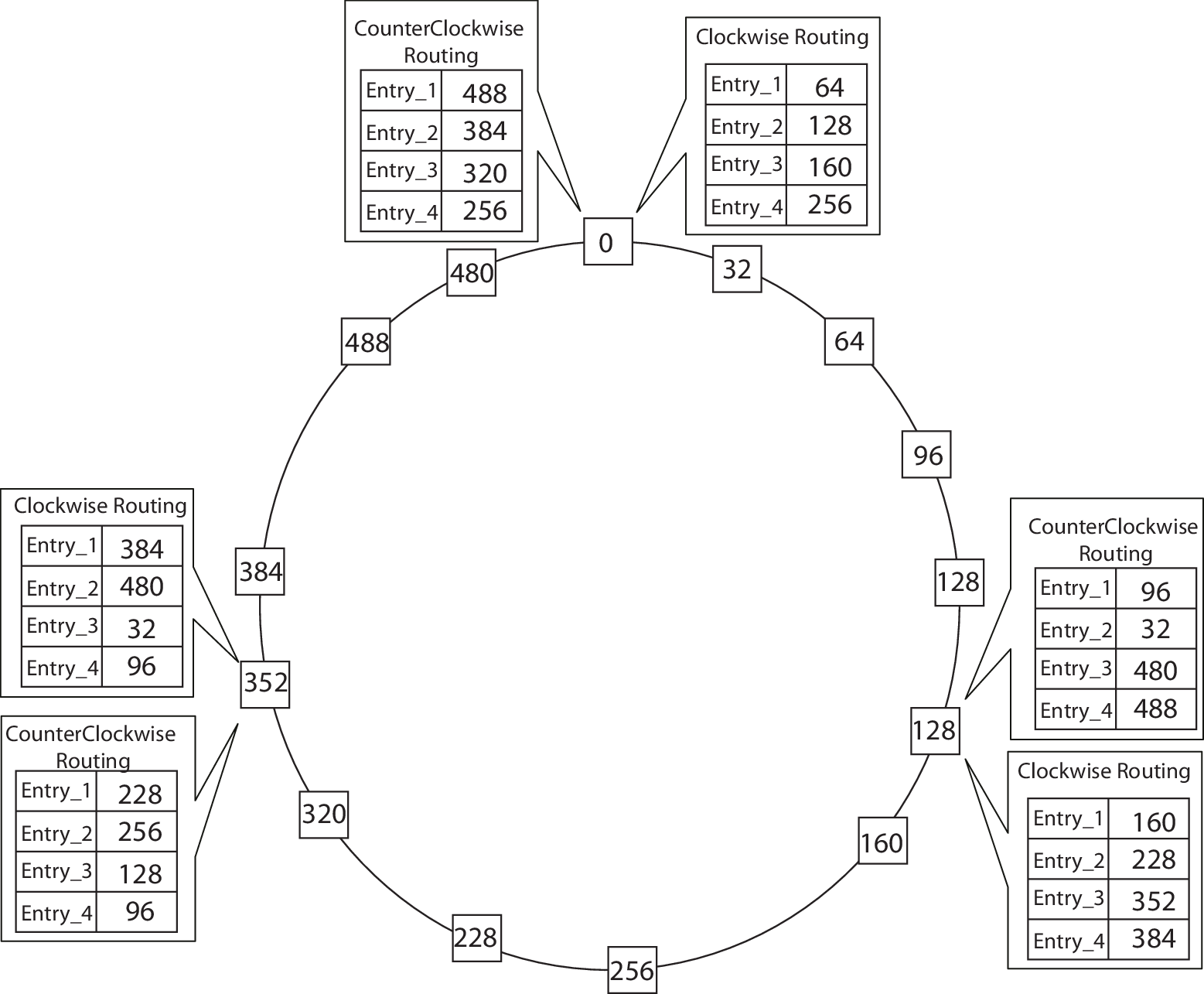}
\caption{An example of bidirectional routing tables for supreme-nodes.}
\label{fig_7}
\end{figure}

As we mentioned, clusters and sections may remain inactive until at least one node of that group joins the network. But this scheme raises a question that what happens to routing mechanism for a supreme-node (explained in section 3.2) when the destination is not active. In this case, the predecessor supreme-node of absent section should take its place instead. To better understand this situation, we present Fig.~\ref{fig_7}.   

The network has a maximum number of $2 ^ 9 = 512$ nodes. This network is divided into 15 sections. Supreme-nodes of 192 and 416 are absent in the network and their sections are inactive. As mentioned before, we emulate the functionality of binary search. To do this, each supreme-node has two routing tables. Using this scheme helps us throw away at least half of the remaining useless nodes in each step. Of course halving the steps is a worst case scenario and it can most probably narrow the scope of searching much better in ordinary cases. 

In the network, N0 shows this design in an ideal state. As we can see, all the destinations for its routing tables are already in the network. It is also apparent that the last entries for both tables are the same. In supreme-node of 128, Entry\_1 has to be filled with 192 but since that section is inactive so it fills with its first predecessor, 160. Another case is the last entries for its routing tables. They both had to point to 416 but since that section is also inactive, each of them are filled with different nodes. The other case is supreme-node of 352 which gives us another example of how this process takes place.

For updating these routing tables, we can use either a periodic or reactive approach. Taking this decision is completely dependent on the network itself. But we should remind that even if we take a periodic approach, its negative effects for overloads should be much smaller than a network such as Chord. Since it is only used for supreme-nodes which are two layers above the regular nodes.

\section{Simulations and analysis}
\label{sec:6}

In a non-P2P environment when there are fixed servers to help the network, everything should work flawlessly. But to truly analyze our method, we need to put it in a stressing situation of P2P where super-nodes leave their positions. In our experiments, we are looking to examine four main aspects of BSROne which are as follows:
 
\begin{itemize}
\item routing efficiency.   
\item overhead over churn
\item fault tolerance
\item stage of stability
\end{itemize} 

In most of our experiments, we used the default mode of BSROne. But by extension, they are also practical to scalable modes. For instance, churn and stage of stability could have the same concepts for both supreme and super-nodes. Therefore, we did not repeat the experiments for each of them. However, in routing efficiency it is only applicable to the supreme-nodes in scalable mode.

\subsection{Routing Efficiency}

In this section, we want to analyze the time that is required for finding nodes in the network. As the name of BSROne suggests and with its methodology, it is obvious that nodes can be found with at most one jump in the default state. But when we are talking about the scalable mode with very large networks where supreme-nodes are used, we actually need to do routing. As a result, we need to experiment on this exact situation. In such networks, it does not matter how many nodes are in the network but how it has been divided into smaller pieces is important. This means that the number of jumps has a direct relation to the number of supreme-nodes in the network.

\begin{figure}[!h]
\centering
\includegraphics[width=72mm]{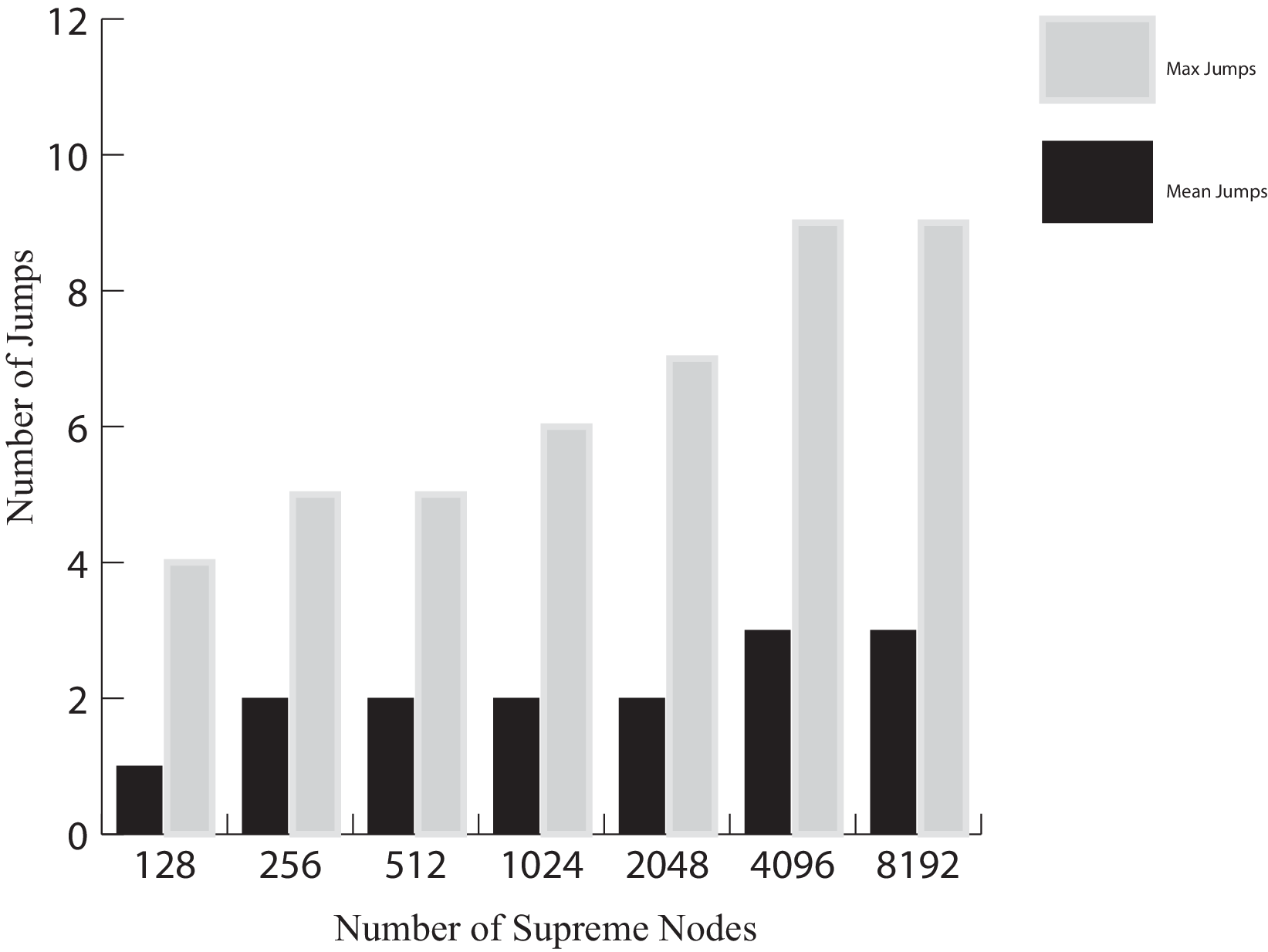}
\caption{The number of mean and maximum jumps in the supreme-nodes.}
\label{fig_8}
\end{figure}

In this experiment, we divided our network into 32 nodes. Each supreme-node was exactly behaving as it was described in section 3.3 and section 5 with two routing tables in clockwise and counterclockwise directions. Each of them was also aware of its successor and predecessor nodes. This is required to fulfill the situation described in section 5. In the network, we generated at least 10 messages between two randomly selected supreme-nodes in each step and measured their jumps. The result is shown in Fig.~\ref{fig_8}. 
 
In this figure, one column represents the mean of jumps and the other is used for maximum number of jumps. Even though the maximum number of jumps had high growth rates in most of the cases but the mean jumps was still growing much more moderately. We have to remind it again that this is not the number of nodes in the network. Since that is completely dependent on how it is going to be managed. A network with 32 nodes for each section can get completely similar results with a network of 1024 nodes in each section.

\subsection{Overhead over Churn}

When new nodes join the network, they either go into a cluster or replace a current super-node. The first case is very simple. The only thing which should be done is to find the right super-node. But for the second case, signals should be sent to all the other super-nodes and inform them about the change. This is very similar to the case that a regular node increases its ability to a point where it is suitable for taking a job of higher position. Therefore, we put them in the same group. 

\begin{figure}[h]
  \centering
  \subfloat[]{\label{first}
    \includegraphics[width=6cm]{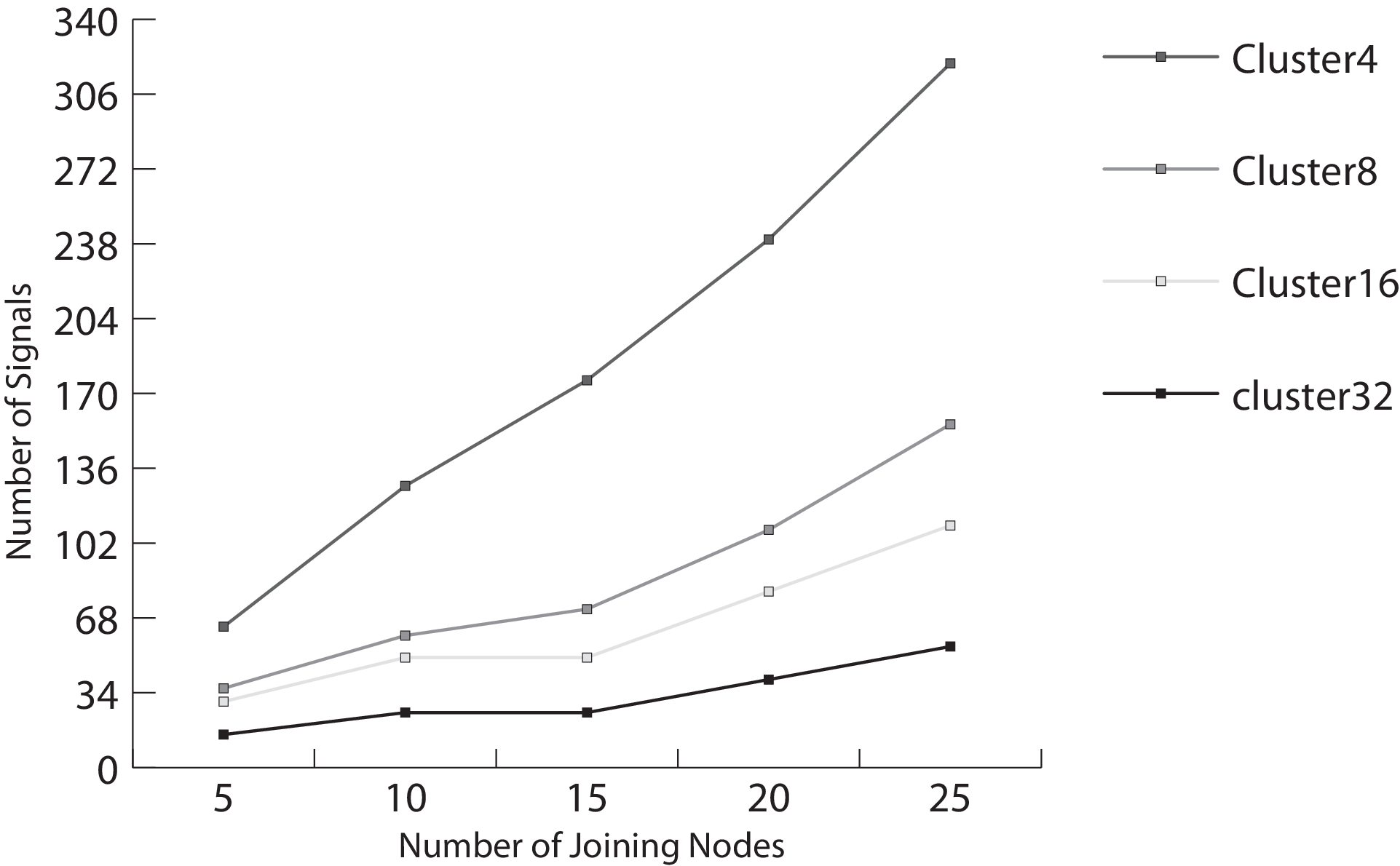}
  }
  \subfloat[]{\label{second}
    \includegraphics[width=6cm]{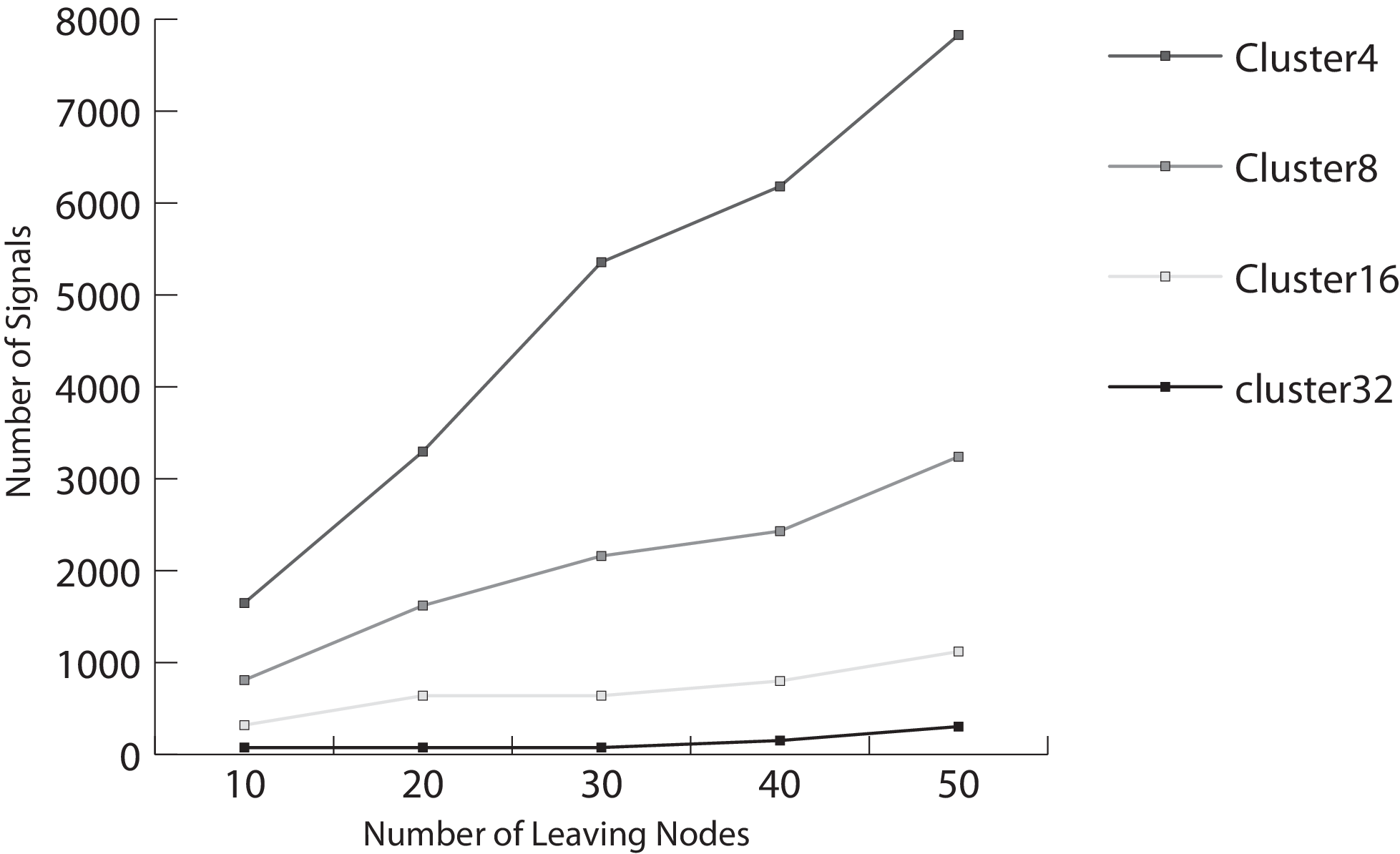}
  }
  \caption{The number of signals when churn happens.}
  \label{fig_9}
\end{figure}

In our first experiment, we want to know how much signals will be sent to the network when new nodes join the network. We made a network of 128 for its maximum number and filled it with 20 nodes which was circulated randomly in the network. In each step, we have increased the number of new nodes to join the network. This process was repeated for several types of clusters. The result is shown in Fig.~\ref{fig_9}.a. As we can see, not only can we get lower signals by using wider cluster range, but also we can attain a much lower growing rate for overlay. This is more obvious when we consider that signals, which are needed for cluster 4, have grown from 64 to 320. But for cluster 32, this range was from 15 to 55.  

To complete our analyses for churn, we also want to analyze the impacts of leaving nodes. When a regular node leaves, the only thing is required for a super-node is to update its searching table. But this is specially important where a super-node leaves the network. In this case, two kinds of signals should be sent. The first is the substitute node which informs the other nodes that it has taken the job. Another signal is for the real super-node which takes the job from the substitute node.

Therefore, what we want to examine is the number of overhead signals for both of these actions. In this case, we used a network with the maximum number of 128 nodes and with initial nodes of 120. As we mentioned before, no signal is needed to be sent for regular nodes but we most probably need at least two signals for each super-node. Nodes were randomly chosen to be removed from the network. The result is shown in Fig.~\ref{fig_9}.b. In this case, we also can see how much cluster management has dramatic effects on the network. The more we choose smaller clusters, the more it will bring overheads on the network. For cluster 32, the overhead has risen from 32 to mere 304. But in cluster 4, overhead has risen from 1648 to 7828. 

\subsection{Fault Tolerance}

One of the most important issues for the network design is fault tolerance. This problem is especially important for structured P2P networks. The reason is that by leaving nodes, a network may lose its connectivity. In the BSROne design, this issue should have a lesser impact since the network is divided into smaller clusters. Therefore, by losing a cluster only one portion of network will be lost and the other parts can function as before. Even though this is considered to be a more flexible approach for this issue, but it is still a very important matter which should be examined carefully. We have realized in the last section that how  bigger clusters can be beneficial for the network. But here it is in complete contrast for this problem where the larger number of nodes may lose their connectivity with bigger clusters.

\begin{figure}[!h]
\centering
\includegraphics[width=62mm]{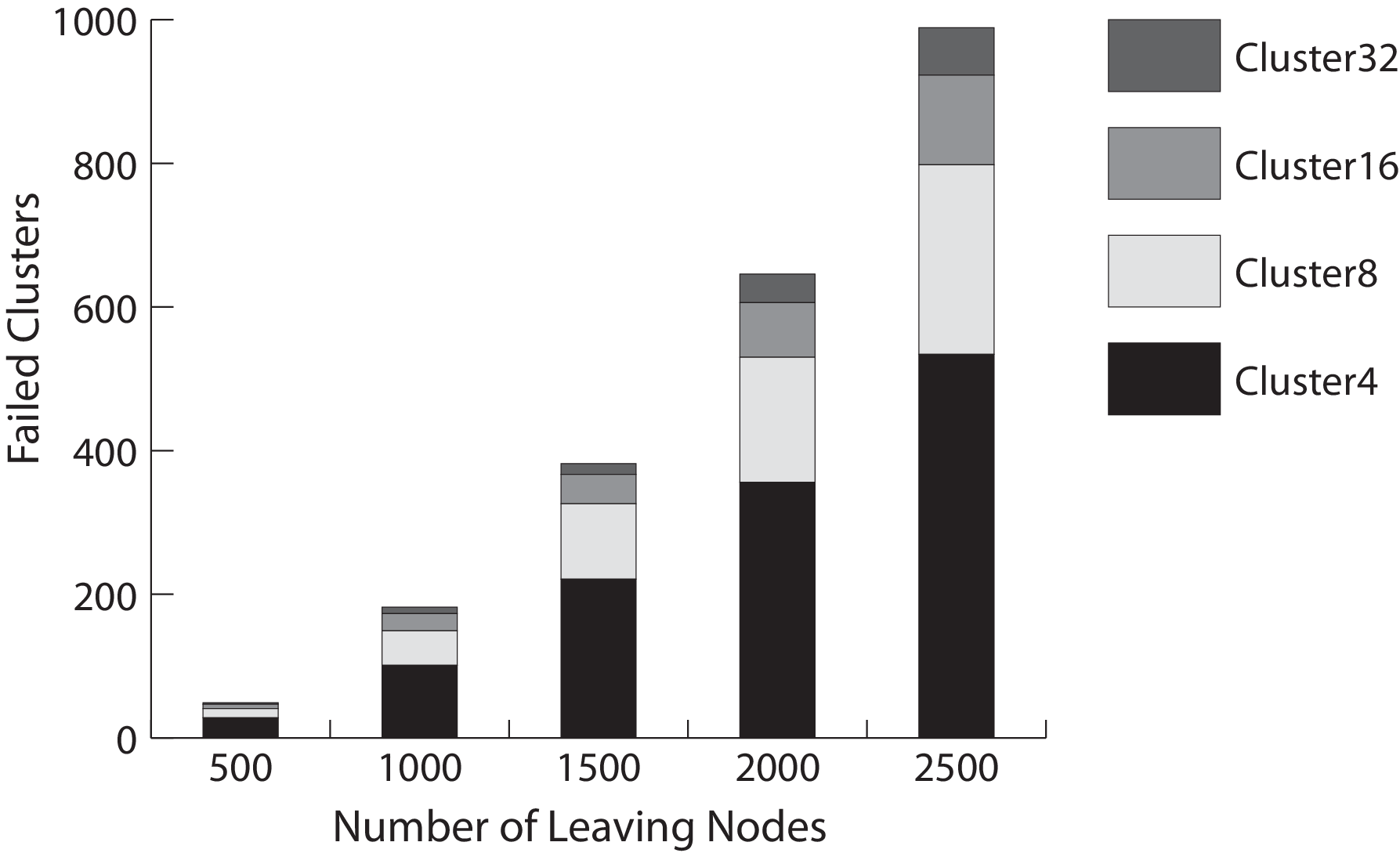}
\caption{The number of failed clusters based on leaving nodes.}
\label{fig_10}
\end{figure}

This time, we used a bigger network with at most 4096 and initially filled all of its positions. Each super-node in each cluster had only one substitute. After a super-node failed, the substitution would immediately replace its place. The result is shown in Fig.~\ref{fig_10}. We have taken snapshots for every 500 leaving nodes and recorded the number of failed clusters based on each cluster type. As it was expected, the more we use smaller clusters, the more we experienced network failures. But in this case, this is not necessarily a bad thing. It is true that there are more failures, but those failures are related to smaller groups. In a network with larger clusters, the failures should be more sever. To remedy this situation, the network managers can define more than one substitute based on their situation. In this way they can enjoy the less overheads that larger clusters bring to the network and still manage fault tolerance properly. 

\subsection{Stage of Stability}

In the last section, we talk about the stages of stability. It is a phenomenon which may happen in such a design. This means that when network is in its nascent state, by churn, a lot of role exchanges may happen. But by passing the time, nodes with better attributes such as bandwidth, stability etc.(as mentioned in section 4.3) take the higher responsibilities. As a result by joining new nodes a role change is less likely to happen. 

\begin{figure}[!h]
\centering
\includegraphics[width=62mm]{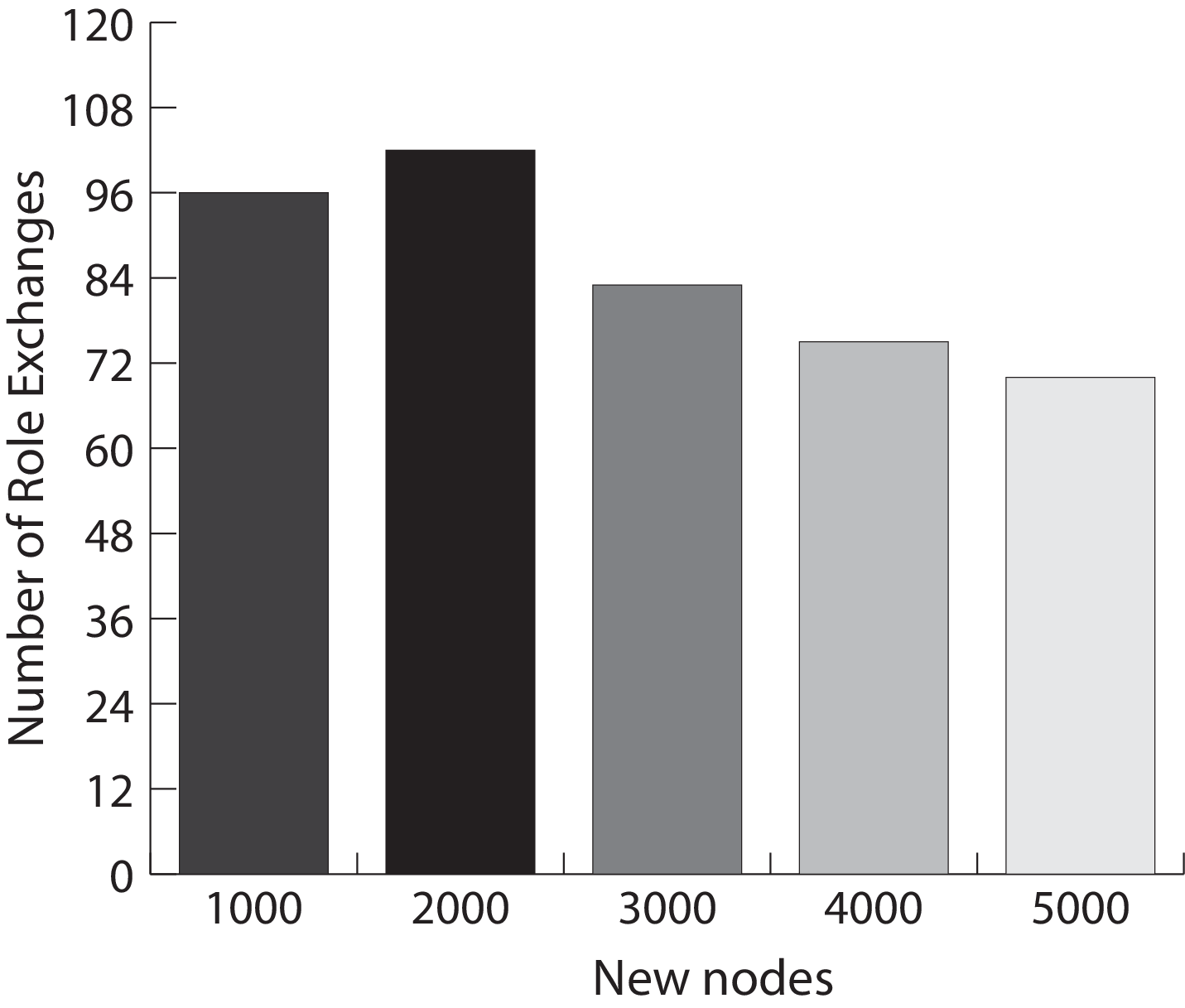}
\caption{The number of role exchanges by arriving new nodes.}
\label{fig_11}
\end{figure}

For this experiment, we used a network of 128 for its maximum nodes and 50 nodes for its initial state. What we have done was to gradually add nodes to the network and then allowed the best nodes to become super-nodes. Leaving nodes were managed in a way that it would match the joining nodes so network could still accept members. To better understand this phenomenon, we present Fig.~\ref{fig_11}. As we can see, the number of the exchanges is shown per thousands of new nodes. It was revealed that after two thousands of nodes, the network went into a stability condition and the number of exchanges has began to decrease. Of course this depends on a lot conditions. The most important attribute for reaching this state is stability and how much managers care for this criteria when choosing super-nodes(see section 4.3).

\section{Conclusion}
\label{sec:7}

An O(1) design for circular structured P2P networks has two important issues: high overload in large networks and process of updating DHT tables. In this paper, we presented BSROne which is specially focused on these two problems. In this attempt, we showed that how it would be possible to achieve O(1) for routing with a reasonable time for updating their tables by clustering the network and using super-nodes. We also demonstrated how such a design could remove the scalability problem with introducing the supreme-nodes and emulating methodology of binary search for them. In addition, we have proposed a criteria selection method which gives a network designer an opportunity to choose between different priorities. In the end, we have examined different aspects of our method with simulations. In that section, we experimented on scalable scenarios and implementations of binary search for the supreme-nodes. We also discussed on the importance of cluster management. It was revealed that by separating network and getting help from a substitute, we can improve fault tolerance of the network significantly. We finally presented an interesting phenomenon which can help the network with its overloads. Overall, we have seen that BSROne is a very effective design which can even surpass its original premises which it was created for.

\bibliography{mybibfile}

\end{document}